\documentstyle[12pt,epsf]{article}

\advance\voffset by -1.5cm
\advance\hoffset by -1.5cm
\def\be{\begin{equation}}
\def\ee{\end{equation}}

\def\pmb#1{\setbox0=\hbox{#1}
 \kern-.025em\copy0\kern-\wd0
 \kern.05em\copy0\kern-\wd0
 \kern-.025em\raise.0433em\box0 }

\def\3{\ss}
\def\sq{\hbox{\rlap{$\sqcap$}$\sqcup$}}
\def\qed{\ifmmode\sq\else{\unskip\nobreak\hfil
\penalty50\hskip1em\null\nobreak\hfil\sq
\parfillskip=0pt\finalhyphendemerits=0\endgraf}\fi}
\def\half {\frac{1}{2}}

\def\bbbz {{\sf Z\!\!Z}}

\def\ss{\bf S}

\def\C{{\cal C}}

\begin{document}

\thispagestyle{empty}
\def\thefootnote{\fnsymbol{footnote}}
\begin{flushright}
  hep-th/9806011\\
  HUTP-98/A027
 \end{flushright}
\vskip 0.5cm

\begin{center}\LARGE
{\bf String Junction Transitions in the Moduli 
Space of N=2 SYM}
\end{center}
\vskip 1.0cm
\begin{center}
{\large  Oren Bergman\footnote{E-mail  address: {\tt
bergman@string.harvard.edu}} and Ansar Fayyazuddin
\footnote{E-mail  address: 
{\tt ansar@curie.harvard.edu}}}

\vskip 0.5 cm
{\it Lyman Laboratory of Physics\\
Harvard University\\
Cambridge, MA 02138}
\end{center}

\vskip 1.5cm

\begin{center}
June 1998
\end{center}

\vskip 1.5cm

\begin{abstract}
The string theory description of BPS states in
D-brane world-volume field theories may undergo transitions
from open strings to string webs, as well as between different
string webs,
as one moves in the field theory moduli space.
These transitions are driven by the string 
creation phenomenon.
We demonstrate such transitions in the D3-brane
realization of $N=2$ $SU(2)$ Super-Yang-Mills theory.
\end{abstract}

\vskip 1.5cm 
\begin{center}
\end{center}

\vfill
\setcounter{footnote}{0}
\def\thefootnote{\arabic{footnote}}
\newpage

\renewcommand{\theequation}{\thesection.\arabic
{equation}}

\section{Introduction}

\noindent In type IIB string theory there exists an 
$SL(2,Z)$ multiplet of strings
labeled by their NSNS and RR 
charges $(p,q)$ \cite{schwarz_pq,witten_bound}.
These strings are allowed to connect at a point or a 
``junction'' as long
as charge is conserved at the junction
\cite{asy,schwarz}.
Of particular interest are
BPS saturated junctions \cite{dasmuk,sen_net},
which preserve a fraction (1/4) of the supersymmetry,
and are therefore guaranteed to be stable. 

Recent interest in string webs has focused on their 
application to quantum field theory 
\cite{gaberdiel_z,bergman,bergman_f,imamura,
michaelov_n_s}.  
One of the successes of
string web technology was the first explicit 
construction of 
$1/4$ BPS states in $N=4$ $SU(N_c)$ Super-Yang-Mills 
(SYM) theory
\cite{bergman}. In this example
the dilaton and axion of type IIB string theory
are constant, and it is relatively straightforward 
to analyze the BPS properties of string webs.
A richer set of examples involve a dilaton and axion
which are position dependent due to the presence 
of 7-branes in the background. 
However, the BPS nature of string webs has
only been analyzed in the 7-brane background
corresponding to pure $N=2$ $SU(2)$ SYM so far
\cite{bergman_f,michaelov_n_s}.
These examples display a number of 
interesting phenomena including ``marginal stability'', 
whereby space 
(moduli space in the world-volume field theory) 
may contain regions in which string webs fail to be BPS.
In this paper we study another interesting phenomenon,
in which the stringy description of a BPS state may
undergo a transition, as one moves in moduli space, 
from either a single open string or a string web, 
to another string web.

In a recent paper we showed how to represent 
BPS states in $N=2$ $SU(2)$ SYM theory 
using string webs \cite{bergman_f}.  
The field theory is constructed as the world 
volume theory of a D3-brane in a 7-brane background.  
BPS states 
correspond to string webs with one 
(or two in the case of vector multiplets) 
string ending
on the D3-brane and the others on the 7-branes.  
Here we show how in different
regions of moduli space the string configuration 
representing
a BPS state changes from a string web to either a
different string web or to a single string.  
These transitions are driven by the string creation
phenomenon 
\cite{bachas_d_g,danielsson_f_k,bergman_g_l,creation}, 
which we generalize here
to multiple string creation between $(p,q)$ 7-branes
and $(r,s)$ strings. This also allows us to generalize
to some extent the ``s-rule'' \cite{hanany_w}, 
which limits the number
of strings that can link a 7-brane and another string.

Transitions of the type addressed here were first
discussed in \cite{gaberdiel_h_z} for the case where 
all strings end on 7-branes. 
A closely related paper \cite{hauer} argues 
that for any given configuration of 7-branes a BPS
state has a unique stringy representative.
 
The results presented here, aside from their application 
to $N=2$ theories,
are of independent interest in string theory.  
In fact,
the 3-brane plays little role in what follows 
except to give strings a place to end.        

The rest of the paper is organized as follows. In
section~2 we review and generalize the string creation
effect and the s-rule. In section~3 we express the 
supersymmetry conditions on strings and string webs in terms 
of $(p,q)$-geodesics and geodesic webs, and relate
the latter to constant phase curves of analytic 
functions. In section~4 we 
show that the complex plane (moduli space) is divided
into four regions, and exhibit the transitions that occur
as one moves from one region to another.
We include an appendix, in which we give relevant
information about the analytic functions in question,
including a few numerically obtained constant phase curves.

\section{String creation and the ``s-rule''}
\setcounter{equation}{0}

\noindent When a D$p$-brane crosses a mutually transverse 
D$(8-p)$-brane a fundamental string, whose ends terminate
on the two D-branes, is created. This phenomenon was 
originally discovered in 
\cite{bachas_d_g,danielsson_f_k,bergman_g_l}, and is 
related by a chain of dualities to the creation
of D3-branes between crossing 5-branes \cite{hanany_w}.
It has since been exhibited in a number of different
ways \cite{creation}.
For $p=1$ the effect involves a D1-brane, {\it i.e.}
$(0,1)$ string, and a D7-brane, {\it i.e.} $(1,0)$ 
7-brane\footnote{$(p,q)$ 7-branes are objects on 
which $(p,q)$
strings can end.}, in type IIB string theory. 
When the branes cross each other, a $(1,0)$
string is created between them. By $SL(2,\bbbz)$ invariance
one can generalize this result to the creation of a single
$(p,q)$ string between a $(p,q)$ 7-brane and an $(r,s)$
string, where $|ps-qr|=1$.

If $|ps-qr|>1$ the configuration cannot be related by
duality to any of the above, and the 
arguments in 
\cite{bachas_d_g,danielsson_f_k,bergman_g_l,creation} 
are not applicable.
Nevertheless, it is straightforward to
generalize the string creation phenomenon to an arbitrary
$(p,q)$ 7-brane and an arbitrary (transverse) $(r,s)$ 
string.
We associate a branch cut to the $(p,q)$ 7-brane in the 
transverse plane, which
one can think of as the place where the following monodromy 
occurs: 
\be
 M_{(p,q)} = \left(
  \begin{array}{cc}
     1-pq & p^2 \\
     -q^2 & 1+pq
  \end{array}
  \right) \; .
\label{monodromy}
\ee  
When an $(r,s)$ string living in the transverse plane
crosses this cut in a counter-clockwise manner 
it undergoes a monodromy transformation,
\be
 \left(
 \begin{array}{c}
   r \\
   s
 \end{array}\right)
 \longrightarrow
 M_{(p,q)}
 \left(
 \begin{array}{c}
   r \\
   s
 \end{array}\right)
 = \left(
 \begin{array}{c}
   r \\
   s
 \end{array}\right)
 + (ps-qr)
 \left(
 \begin{array}{c}
   p \\
   q
 \end{array}\right) \; .
\ee
The asymptotic charges of the string are therefore 
$(r,s)$ at one end, and $(r,s) + (ps-qr)(p,q)$ at the
other (fig.~1a). These cannot change as we pull the string 
through the 7-brane. However, since now the string does
not cross the cut, something else must supply the 
additional charge. We conclude then that $|ps-qr|$
$(p,q)$ strings are created between the string
and the 7-brane, oriented in such a way as to conserve 
the charges (fig.~1b).
\begin{figure}[htb]
\epsfxsize=4 in
\centerline{\epsffile{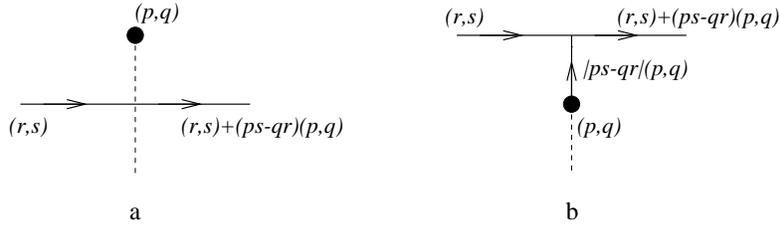}}
\caption{Multiple string creation in $(p,q)$ 7-brane --
$(r,s)$ string background.}
\end{figure}

In the special case of $|ps-qr|=1$ we obtain the
known result of single string creation. 
For $|ps-qr|>1$ one can use $SL(2,\bbbz)$
to relate the system to a $(1,0)$ 7-brane and an $(r',s')$
string. One can then think of the $(r',s')$ string as
a bound state of $|s'|$ D-strings with $r'$ units of
electric flux. Each of the D-strings will create
a $(1,0)$ string when it crosses the 7-brane, giving
a total of $|s'|=|ps-qr|$ strings.

The orientation of the created strings is determined by 
charge conservation, but is also restricted by
supersymmetry. Consider the case of D-branes.
The initial configuration, consisting of a D7-brane
and a transverse D-string, is supersymmetric,
and therefore the final
configuration must also be supersymmetric. 
On the other hand it can be shown \cite{bergman_g_l}
that such a configuration will only be supersymmetric
if the fundamental string is oriented in the right way.
It is a matter of convention which of the two orientations
preserves supersymmetry. The important point is that only
one of them does. Charge conservation in the string 
creation process fixes this convention, and gives us
the unique supersymmetric orientation.

This leads to a non-trivial selection rule, known as
the ``s-rule'' \cite{hanany_w}, 
for configurations of 7-branes and transverse
strings, and configurations related to them by dualities.
Consider a configuration of $n$ fundamental strings between
a D7-brane and a transverse D-string (fig.~2a). 
Supersymmetry restricts 
the orientation of the fundamental strings. As the D-string
crosses the D7-brane two things happen: the $n$ fundamental
strings reverse orientation, and a single fundamental string
oriented supersymmetrically is created. The latter will 
annihilate with one of the former, leaving $n-1$ strings
which are oriented so as to break supersymmetry (fig.~2b). 
The final
configuration is therefore {\it not} supersymmetric, and
therefore neither is the initial configuration, even though
it seemed to be. We conclude that supersymmetry restricts
not only the orientation of strings between the branes,
but also their number. In the above case there can be at
most {\it one} fundamental string between a D7-brane 
and a D-string.
More generally, the maximum number of $(p,q)$ strings between
a $(p,q)$ 7-brane and an $(r,s)$ string allowed by
supersymmetry is $|ps-qr|$.
\begin{figure}[htb]
\epsfxsize=4 in
\centerline{\epsffile{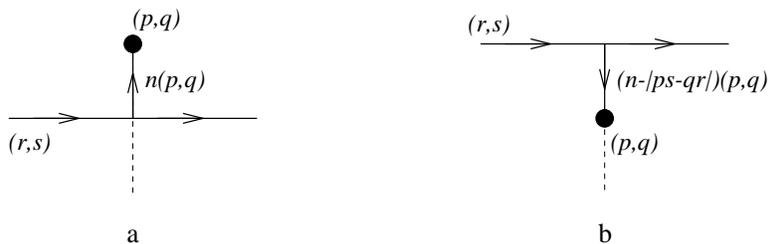}}
\caption{The
configuration in (a) appears to be supersymmetric,
but is not if $n>|ps-qr|$, as can be seen from
the configuration in (b), which is obtained by 
crossing the branes.}
\end{figure}

In the case of $|ps-qr|=1$ there are other ways of deriving
this rule \cite{bachas_green}. In particular, in the dual
configuration of a D0-brane and a D8-brane the ``s-rule''
can be understood as a Pauli exclusion principle, since
the ground state of a 0-8 string corresponds to a single
fermion in the D0-brane world-volume. Alternatively, in 
the D2$\perp$D6 configuration one can derive the ``s-rule''
by lifting to M-theory, and finding all the holomorphic
curves corresponding to a supersymmetric embedding of the
M2-brane in a Kaluza-Klein monopole background. 

As an application of this selection rule let us consider
a configuration of a $(1,0)$ 7-brane, a $(2,1)$
7-brane, and a $(0,1)$ 7-brane, connected by a ``3-string
junction'' consisting of two $(1,0)$ strings, a $(2,1)$
string, and a $(0,1)$ string (fig.~3b). Since the $(2,1)$
and $(0,1)$ 7-branes can be thought of as the quantum
resolution of an orientifold 7-plane \cite{sen_f}, this
configuration corresponds to the quantum resolution of
a string which begins on the D7-brane, reflects off the
orientifold plane, and ends on the D7-brane (fig.~3a).
(In the covering space this becomes a string between
the D7-brane and its image). In this {\it classical}
description, the usual massless (vector) ground state 
of the string
is removed by the orientifold projection, leaving a massive
{\it non-BPS} ground state. 
This state is extremal in the
sense that it is the lowest mass charged state.

In the quantum picture this is understood as follows.
The configuration in (fig.~3b) violates the ``s-rule'',
since it contains {\it two} $(1,0)$ strings between the 
D7-brane and the $(0,1)$ string. It therefore breaks
supersymmetry, and the corresponding state is non-BPS.

One can compare this to the situation in which the D7-brane
is replaced by a D3-brane. In this case the massless
vector 
survives the orientifold projection, and leads to an
enhanced gauge symmetry $Sp(1)\sim SU(2)$ when the D3-brane
coincides with the orientifold 7-plane. In the quantum 
picture the ``3-string junction'' with the D3-brane 
satisfies the s-rule,
and is therefore supersymmetric.  
\begin{figure}[htb]
\epsfxsize=3 in
\centerline{\epsffile{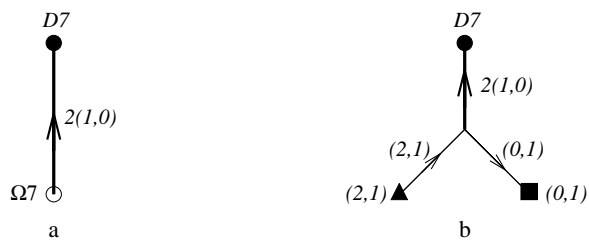}}
\caption{The supersymmetric ground state is removed from 
the spectrum by the orientifold projection in
the classical picture (a), and by the s-rule in
the quantum picture (b).}
\end{figure}

\section{Geodesics and geodesic webs}
\setcounter{equation}{0}

\noindent A $(p,q)$-geodesic between points $A$ and $B$
is defined as the path which minimizes the mass of a 
$(p,q)$ string stretched between $A$ and $B$,
\begin{equation}
m_{(p,q)}= \int_A^B T_{(p,q)}ds \; ,
\end{equation}  
where $T_{(p,q)}= (\mbox{Im}\tau)^{-1/2}|p+q\tau|$. 
Since the geodesic is independent of the overall sign
of $(p,q)$ we shall assume, without loss of generality,
that $p\geq 0$.
In a generic 7-brane background $\tau$ is 
position-dependent,
and in fact undergoes monodromies as branch cuts are 
crossed. The metric on
the transverse (complex) plane is given by \cite{vafa}
\be
  ds^2 = \mbox{Im}\tau|f(z)dz|^2 \; ,
\ee 
and is smooth everywhere. The function $f(z)$ is determined
by the 7-brane background.
One can define a flat ``$(p,q)$-metric''
\be 
  ds_{(p,q)} = T_{(p,q)}ds = |(p+q\tau)f(z)dz| \; ,
\label{pqmetric}
\ee
and thereby coordinates $w_{(p,q)}$ given by
\be
 dw_{(p,q)} = (p+q\tau)f(z)dz \; ,
\label{pqcoord}
\ee 
in which $(p,q)$-geodesics are straight lines,
\be
 w_{(p,q)}(t) = ct + d \; .
\ee
A necessary condition for $(p,q)$ strings to
preserve supersymmetry is that they lie along
$(p,q)$ geodesics \cite{sen_bps}. If in addition
the 7-brane background is supersymmetric, this is
a sufficient condition.

Multi-pronged strings, and more generally 
string webs (which can include several junctions and
internal strings), arise quite naturally in type IIB
string theory. The simplest example corresponds to the
configuration of a fundamental string ending on a 
D-string. By charge conservation, this turns the D-string
into a $(1,1)$  string, and one ends up with a 
$((1,0),(0,1),(1,1))$ three-pronged string.
Due to the $SL(2,\bbbz)$ symmetry of type IIB string
theory there is actually an entire $SL(2,\bbbz)$
multiplet of such objects, each of which satisfies
the charge conservation condition $\sum p_i=\sum q_i =0$.
By gluing several of these objects together one can form
a rich variety of string webs.

A necessary condition for a string web to 
preserve supersymmetry
is that it lies on a {\it geodesic web}.
Geodesic webs consist of several connected 
$(p,q)$-geodesics,
\be
 w_{(p_i,q_i)}(t_i) = c_it_i + d_i \;,
\ee
such that the phases of $c_i$ are the same for all
$i$, {\it i.e.}
\be
 c_i = e^{i\phi} |c_i| \;.
\label{slopes}
\ee
This reduces to the usual angle conditions in
flat space \cite{dasmuk,sen_net}.
This is not a sufficient condition however,
as the string web may violate the ``s-rule''
(or a generalization of it given in
\cite{michaelov_n_s}), as in the example of the previous
section.

The background of interest consists of a 
$(0,1)$ 7-brane at $z=1$ and a $(2,1)$ 7-brane 
at $z=-1$. 
The metric on the complex plane 
is given by
\be
 ds^2 = 2\mbox{Im}\tau\left|
  \eta^2(\tau)(z-1)^{-1/12}(z+1)^{-1/12}dz\right|^2 \; ,
\ee
where we have set $8\Lambda^2 = 1$ as compared to 
\cite{bergman_f}.
We shall choose the branch cuts associated to the
$(0,1)$ and $(2,1)$ 7-branes to run along
the real axis from $z=1$ to $z=-\infty$,
and from $z=-1$ to $z=-\infty$, respectively.
The monodromies associated with the cuts along 
$[-\infty,-1]$ and $[-1,1]$ are respectively
given by
\be
 M_\infty = \left(
 \begin{array}{rr}
  -1 & 4 \\
   0 & -1 
 \end{array}\right) \; , \qquad 
 M_{(0,1)} = \left(
 \begin{array}{rr}
   1 & 0 \\
   -1 & 1 
 \end{array}\right) \; .
\ee
As a result, the charge of the 7-brane at $z=-1$
is $(2,1)$ when viewed from above the cut, and 
$(2,-1)$ when viewed from below the cut.

The above metric is related
to the solution of $N=2$ $SU(2)$ SYM \cite{SW} 
as follows \cite{sen_f}:
\be 
 T_{(p,q)} ds = |pda + qda_D| \; ,
\label{swconnection}
\ee
where $a(z)$ and $a_D(z)$ are given by integrals of
the meromorphic Seiberg-Witten differential over the two cycles of
the auxiliary Riemann surface. 
These can be expressed in terms of hypergeometric 
functions
(eq.~(\ref{hypergeo}), appendix).
Comparing with (\ref{pqmetric}) and (\ref{pqcoord})
we see that the coordinates
in which a $(p,q)$ geodesic is straight are given by
\be
 w_{(p,q)} = pa(z) + qa_D(z) \; .
\ee
The connection with $N=2$ SYM is not a coincidence, 
and in fact
follows by considering a D3-brane probe in the above
background \cite{banks_d_s}. 
In particular, the BPS states 
are realized as strings ending on the D3-brane
\cite{sen_bps}.
By supersymmetry, these strings should
lie on geodesics given by 
\be
 pa(z(t)) + qa_D(z(t)) = ct + d \; ,
\label{gensolution}
\ee
where $z(0)$ is the position of the D3-brane.
Note however that $(p,q)$ and $(a,a_D)$ undergo 
monodromies
as one crosses branch cuts. We therefore define smooth
multi-valued functions $\tilde{a}(z)$ and 
$\tilde{a}_D(z)$ by analytic continuation.
In terms of these, the geodesic equation 
(\ref{gensolution}) holds across cuts as well. 

The BPS spectrum of $N=2$ $SU(2)$ SYM in the
weakly coupled regime consists of $W$-boson
vector multiplets carrying charges $(\pm 2,0)$, and
dyon hypermultiplets carrying charges $(2n,\pm 1)$.
The corresponding geodesics would have to be given 
by:
\begin{enumerate}
 \item A $(\pm 1,0)$-geodesic that begins
  on the D3-brane, goes around the two 7-branes, 
  and ends on the D3-brane,
 \be 
  2\tilde{a}(z(t)) = c(t-1/2) \; .
\label{w_geodesic}
\ee
\item $(2n,\pm 1)$-geodesics that begin on the D3-brane
  and end on one of the 7-branes,
\be
   2n\tilde{a}(z(t)) \pm \tilde{a}_{D}(z(t)) = c(t-1) \; .
\label{hyper_geodesic}
\ee
\end{enumerate}
In both cases there is a single complex parameter $c$.
These geodesics are therefore given by {\it constant 
phase curves} (CPC) of the appropriate analytic function
appearing on the left hand side of (\ref{w_geodesic})
and (\ref{hyper_geodesic}).
In order to end on either 7-brane, a geodesic of the
second kind with 
$n>1$ would have to cross the cut along $[-\infty,-1]$
(and thereby undergo the monodromy $M_\infty$)
$n/2$ times if $n$ is even, and $(n-1)/2$ times if 
$n$ is odd, without ever crossing the cut along 
$[-1,1]$ (where it would undergo the monodromy $M_{(0,1)}$).

Alternatively, the same states (with the exception
of $n=0,1$ in the second case) could also be described
by string webs connecting all three branes, and
lying on geodesic webs consisting of three
geodesics \cite{bergman_f}. If we denote the positions
of the D3-brane and junction by $z_{3}$ and $z_0$ 
respectively, the appropriate webs are given by
\begin{eqnarray}
 \label{wweb}  
\tilde{a}_D(z(t_1)) &=& \tilde{a}_D(z_0)t_1 \nonumber\\
 \mbox{}[2\tilde{a}+\tilde{a}_D](z(t_2)) &=& 
       [2\tilde{a}+\tilde{a}_D](z_0)t_2  \\
\tilde{a}(z(t_3)) &=& 
           \Big(\tilde{a}(z_0) - a(z_3)\Big)t_3 
          + a(z_3)\nonumber
\end{eqnarray} 
for the $W$-boson, and
\begin{eqnarray}
 \label{dyonweb}
 \tilde{a}_D(z(t_1)) &=& \tilde{a}_D(z_0)t_1 \nonumber\\
 \mbox{} [2\tilde{a}+\tilde{a}_D](z(t_2)) &=& 
       [2\tilde{a}+\tilde{a}_D](z_0)t_2 \\
 \mbox{}[2n\tilde{a}\pm\tilde{a}_{D}](z(t_3)) &=& 
  \Big([2n\tilde{a}\pm \tilde{a}_D](z_0)
  - [2na\pm a_D](z_3)\Big)t_3 
  + [2na\pm a_{D}](z_3)\nonumber
\end{eqnarray} 
for the dyons. In the above we have assumed that the
geodesics begin on the branes and end at the junction, so
$t_i=0$ corresponds to the position of the 
$(0,1)$ 7-brane, $(2,1)$ 7-brane, and D3-brane
for $i=1,2,3$ respectively,
and $t_i=1$ to the position of the junction. 
Consequently, we can drop the tilde from the functions
evaluated at the positions of the branes.
Note that the equations are independent of the 
positions of the 7-branes, since the corresponding
functions $pa+qa_D$ vanish there.
The condition on the slopes (\ref{slopes}) is
satisfied provided that \cite{bergman_f}
\be
 \mbox{Im}{\tilde{a}_D(z_0)\over \tilde{a}(z_0)} = 0\;, \qquad
 {\tilde{a}_D(z_0)\over \tilde{a}(z_0)} > -2 \; ,
\label{condition1}
\ee
for both the $W$-boson and the dyons, and
\be
 \mbox{Im}{a(z_{3})\over \tilde{a}(z_0)} = 0 \;, \qquad
 {a(z_{3})\over \tilde{a}(z_0)} > 1 
\label{wcondition2}
\ee
for the $W$-boson, and
\be
 \mbox{Im}{2na(z_{3})\pm a_D(z_{3})\over
   2n\tilde{a}(z_0)\pm \tilde{a}_D(z_0)} = 0 \;, \qquad
   {2na(z_3)\pm a_D(z_3)\over 2n\tilde{a}(z_0)
     \pm \tilde{a}_D(z_0)} > 1 \; ,
\label{dyoncondition2}
\ee
for the dyons.
The conditions in (\ref{condition1})
imply that the junction ($z_0$) must
lie on the curve of marginal stability 
$\C_M$ (see appendix for definition of $\C_M$).
The first condition in (\ref{wcondition2}),
implies that the third
geodesic in (\ref{wweb}) 
{\it i.e.} the one beginning at the D3-brane ($z_3$)
and ending at the junction ($z_0$),
is a CPC of the function $\tilde{a}(z)$.
Similarly, the first condition in (\ref{dyoncondition2}) 
implies that
the third geodesic in (\ref{dyonweb}) is a CPC
of the function $2n\tilde{a}(z)\pm \tilde{a}_D(z)$.
The second condition in both (\ref{wcondition2})
and (\ref{dyoncondition2}) implies that 
the D3-brane must lie {\it outside} $\C_M$ 
(see appendix B of \cite{bergman_f}).
Since the the other two geodesics begin on
7-branes, they too are CPC's,
and we conclude that the entire geodesic web 
consists of three CPC's of equal phase,
which begin at the location of the branes
and end at a common point on $\C_M$.

\section{Transitions}
\setcounter{equation}{0}

\noindent Uniqueness of the BPS state demands that
there be only one string configuration which
describes it at a given point in moduli space.
This can either be a single string on a smooth
geodesic, or a string web on a geodesic web.
As the above geodesics and geodesic webs are 
associated with CPC's of analytic
functions, one can determine their existence by
studying these curves.
Specifically, one is interested in CPC's of the functions
$\tilde{a}(z)$ and $2n\tilde{a}(z) \pm \tilde{a}_{D}(z)$.
These can be determined numerically from the
explicit expressions (\ref{hypergeo}) for $a, a_D$.
We have plotted CPC's of the
single-valued functions $a(z)$ and  
$2n{a}(z) \pm {a}_{D}(z)$ for a few values
of $n$ using Mathematica (fig.~6, appendix). The CPC's 
of the corresponding smooth functions 
$\tilde{a}(z)$ and 
$2n\tilde{a}(z) \pm \tilde{a}_{D}(z)$ are obtained 
by gluing those of the appropriate single-valued
functions at the cuts. 

 From figure~6 it is apparent that all the CPC's of $a(z)$ 
intersect $\C_M$, and then proceed to intersect the cut
along $[-1,1]$. A closed geodesic which goes around 
both 7-branes therefore does not exist.
This has also been shown in a different way in 
\cite{gaberdiel_h_z} and in \cite{bergman_f}. 
On the other hand, since all the CPC's cross $\C_M$
before crossing any cuts, there exists a
geodesic web of the form (\ref{wweb}) 
everywhere {\it outside} $\C_M$ \cite{bergman_f}.
This is consistent with the observation that the $W$-boson
is stable everywhere outside $\C_M$.

The CPC's of $a_D(z)$ and
$2a(z)\pm a_D(z)$, {\it i.e.} $n=0,1$, 
all end on the corresponding 7-brane.
This means that
there exists a geodesic starting at any point in 
the complex plane and ending on the appropriate 
7-brane. This is consistent with the observation
that the corresponding BPS states exist everywhere
in moduli space.

For $n\geq 2$ the moduli space is divided 
according to the behavior of the CPC's into
four distinct regions, which are separated by $\C_M$, the
cut along $[-\infty,-1]$, and the two CPC's corresponding
to $\phi=0$ and $\phi=\pi/2$ (fig.~4).
\begin{itemize}
\item In region I (inside $\C_M$) CPC's intersect the cut along
$[-1,1]$. As this would imply a monodromy $M_{(0,1)}$,
$(2n,\pm 1)$-geodesics with $n\geq 2$ do not exist in 
this region (this has also been shown in \cite{ansar}).
In addition, geodesic webs can only exist when the D3-brane
is {\it outside} $\C_M$, so they too do not exist in region I.
This is consistent with the observation that there are no
BPS dyons with $n\geq 2$ inside $\C_M$.
\item In regions II and III CPC's intersect the lower half 
and upper half of $\C_M$, respectively.
They then proceed to intersect the 
cut along $[-1,1]$. Smooth geodesics therefore do
not exist.
In these regions the BPS states
are represented by string webs with a single junction
point located on $\C_M$. Since the charge
of the 7-brane at $z=-1$ is different in the upper
and lower half planes, the appropriate string webs
will be different in regions II and III as well.
\item In region IV CPC's intersect the cut along 
$[-\infty,-1]$. The string therefore undergoes 
the monodromy $M_\infty$. For $n=2$ it becomes 
a $(0,1)$ string which then follows a $(0,1)$ geodesic
ending on the $(0,1)$ 7-brane at $z=1$. 
For $n>2$ the $(2n,\pm 1)$
geodesic in region IV becomes a $(2n-4,\pm 1)$
geodesic in region II, which in turn intersects
the lower half of $\C_M$. These states are therefore
represented by string webs.
\end{itemize}

As the D3-brane moves from one region to another
one expects transitions between different string
webs which represent the same BPS state. In the particular
case of the $(4,1)$ dyon we expect a transition 
from a single string to a 
$((4,1),2(2,1),(0,1))$
string web as the D3-brane
crosses the $\phi=\pi/2$ contour, and a transition
from a $((4,1),2(2,1),(0,1))$ web to a 
$((4,1),2(2,-1),3(0,1))$ web as it crosses the $\phi=0$
contour (figure 5). From the discussion in section~2
we see that this is precisely what happens.
When the $(4,1)$ string crosses the $(2,1)$ 7-brane at
$z=-1$ {\it two} $(2,1)$ strings are created, leading to
the first transition. As the D3-brane moves towards
the $\phi=0$ contour the junction point slides along $\C_M$,
until it coincides with the $(0,1)$ 7-brane at $z=1$ when
the D3-brane is on the $\phi=0$ contour. At this point 
the $(4,1)$ string crosses the $(0,1)$ 7-brane, thereby
creating {\it four} $(0,1)$ strings. One of these annihilates
with the original $(0,1)$ string which has reversed orientation,
leaving the required {\it three}. 
This is the second transition.

\newpage

\begin{figure}[htb]
\epsfxsize=2.4 in
\centerline{\epsffile{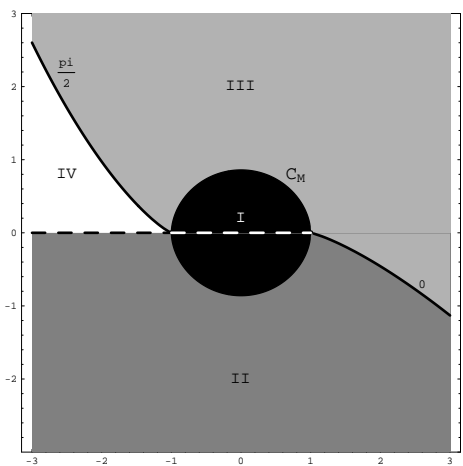}}
\caption{Phase diagram for $(2n,1)$ dyons with $n\geq2$.}
\end{figure}

\begin{figure}[htb]
\centerline{\epsfxsize=2.4 in
\epsffile{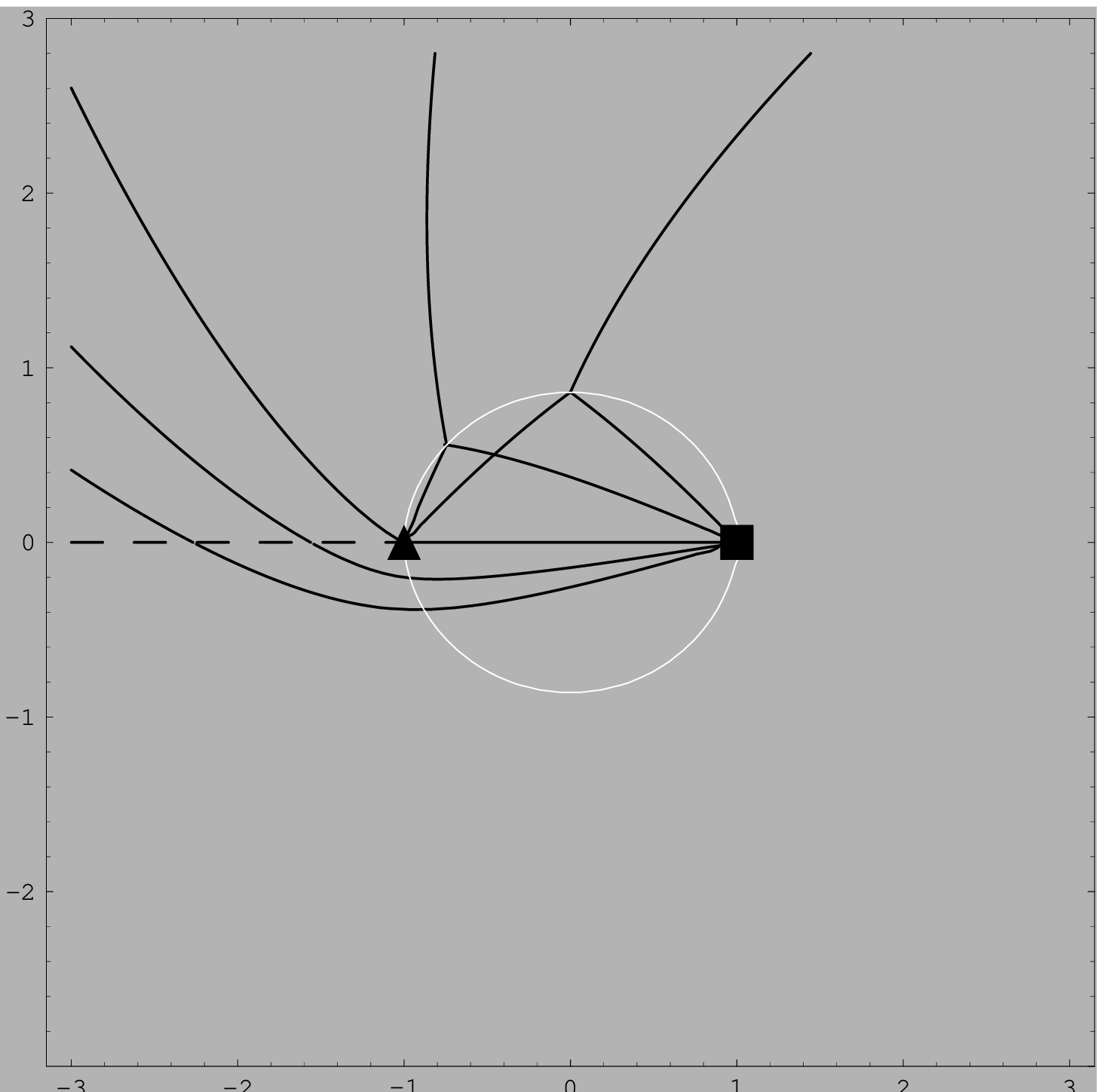}\hspace{10 pt}
\epsfxsize=2.4 in
\epsffile{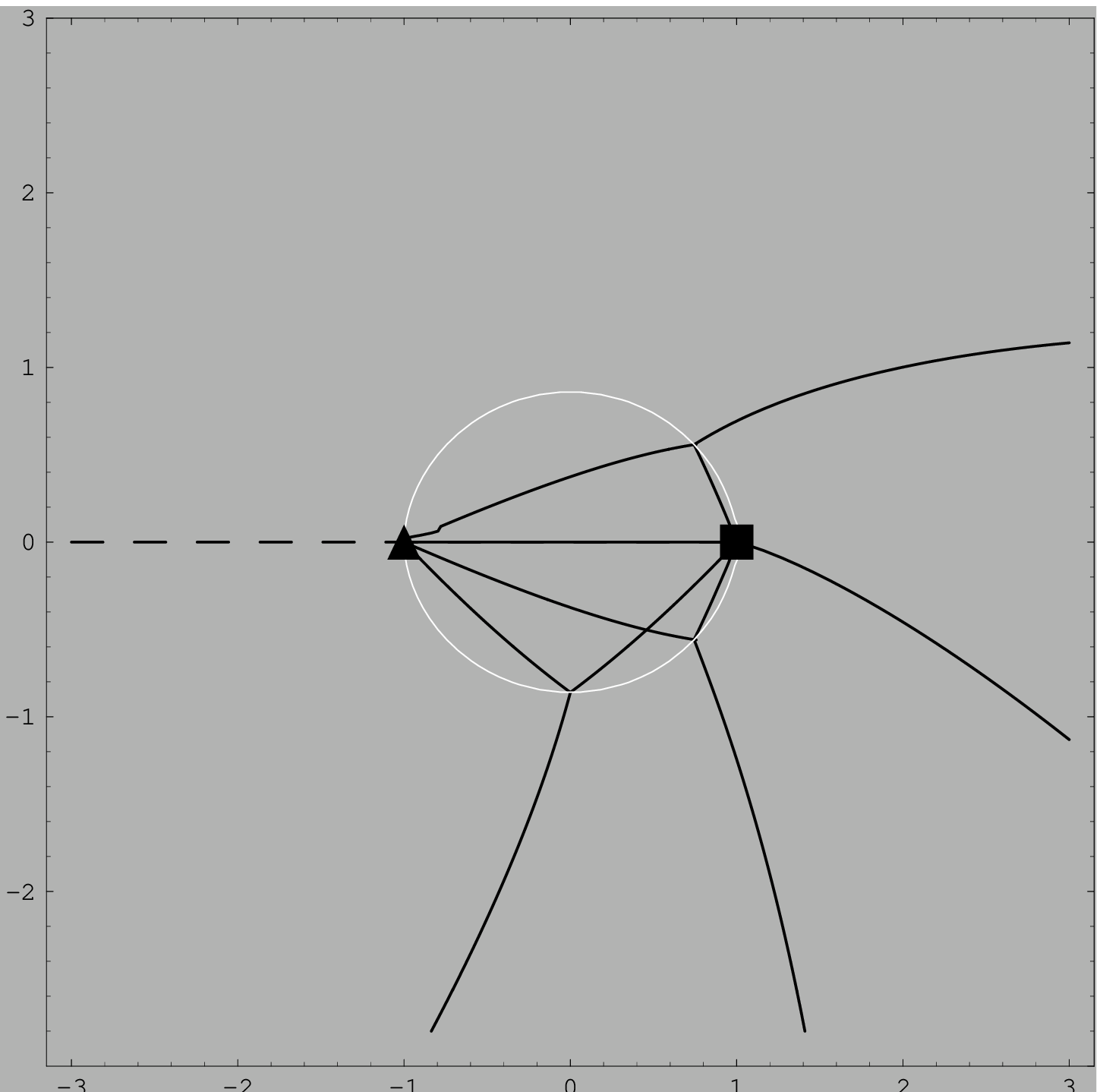}}
\caption{String $\rightarrow$ Web and Web $\rightarrow$
Web transitions for the $(4,1)$ dyon.}
\end{figure}
One can easily generalize this discussion for all the 
$(2n,\pm 1)$ dyons with $n\geq 2$. In each case the string
description undergoes two transitions at $\phi=0, \pi/2$,
which are associated with crossing the two 7-branes.

\section*{Acknowledgments}
This work is supported in part by the NSF under grant 
PHY-92-18167.

\appendix
\section*{Appendix: Relevant functions and curves}
\renewcommand{\theequation}{A.\arabic{equation}}
\setcounter{equation}{0}

\noindent The functions $a(z)$ and $a_D(z)$ can be expressed
in terms of hypergeometric functions as
\begin{eqnarray}
 a(z)& =& \left({z+1\over 2}\right)^{1/2}
     F\left(-\half,\half,1;{2\over z+1}\right)
    \nonumber\\
 a_D(z)&=&  i\left({z-1\over 2}\right)
    F\left(\half,\half,2;{1-z\over 2}\right)\; .
\label{hypergeo}
\end{eqnarray}
We choose the branch cuts (dashed lines in the figures)
to run along the real axis
from $z=-1$ to $z=-\infty$ for $a(z)$, and from 
$z=1$ to $z=-\infty$ for $a_D(z)$.  
The values of the functions at the branch points
$z=\pm 1$ are given by
\be
 a(1)=2/\pi,\;\; a(-1)=\pm 2i/\pi,\;\;
a_D(1)=0,\;\; a_D(-1) = -4i/\pi \; ,
\label{values}
\ee
where the sign in $a(-1)$ depends on whether one is
just above the cut, or just below it.
These points lie on the {\it curve of marginal stability},
defined by
\be
 \C_M : \qquad \mbox{Im}{a_D(z)\over a(z)}=0 \; .
\ee
This curve is diffeomorphic to a circle centered at the
origin. Denote the upper and lower halves by $\C^+_M$
and $\C^-_M$, respectively. It then follows from
(\ref{values}) that
\be
 \begin{array}{rrclcc}
     \mbox{On}\;\; \C_M^+ : \quad 
     &  -\pi & \leq \mbox{Arg}(a_D) = & 
     \mbox{Arg}(a) - \pi & \leq & -\pi/2 \\[5pt]
    \mbox{On}\;\; \C_M^- : \quad 
    & -\pi/2 & \leq \mbox{Arg}(a_D) = & 
     \mbox{Arg}(a) &\leq & 0 \; .  
 \end{array} 
\ee
Consider functions of the form $f(z)=pa(z)+qa_D(z)$.
We can assume, without loss of generality, that $p\geq 0$.
For $p>0$ we find
\be
 \begin{array}{rl}
  \mbox{On}\;\;\C_M^+ : \quad & \left\{
  \begin{array}{rcll}
   -\pi/2 & \leq \mbox{Arg}(f) \leq & 0 & 
  \quad\mbox{if}\;\;\; p<2q\\
   0 & \leq \mbox{Arg}(f) \leq & \pi/2 & 
  \quad\mbox{if}\;\;\; p\geq 2q
  \end{array}\right. \\[10pt]
  \mbox{On}\;\;\C_M^- : \quad & \left\{
  \begin{array}{rcll}
   0 & \leq \mbox{Arg}(f) \leq & \pi/2 & 
   \quad\mbox{if}\;\;\; p<2q\\
   -\pi/2 & \leq \mbox{Arg}(f) \leq & 0 & 
   \quad\mbox{if}\;\;\; p\geq 2q
  \end{array}\right. 
 \end{array}
\ee
Constant phase curves (CPC) are defined by
\be 
  \mbox{Arg}(f(z)) = \phi \; .
\ee
If $\phi$ is in the above ranges the curve intersects
$\C_M$ and ends on the cut along $[-1,1]$. Otherwise it
ends on the cut along $[-\infty,-1]$ without intersecting
$\C_M$.     
When we consider the smooth continuation of $f(z)$
using $\tilde{a}$ and $\tilde{a}_D$, the CPC's continue
across the cuts. In terms of $a,a_D$, a CPC of 
the function $pa(z)+qa_D(z)$ connects smoothly at the cut 
to a 
CPC of a function $p'a(z)+q'a_D(z)$, where $(p',q')$
are related to $(p,q)$ by the appropriate monodromy.

The shapes of the CPC's can be qualitatively deduced
from the asymptotics of $a$ and $a_D$,
\begin{eqnarray}
 a(z)&\sim& \sqrt{z/2}\nonumber \\
 a_D(z) &\sim& {i\over\pi}\sqrt{2z}
    \Big[\ln{z} + 3\ln{2} -2\Big] \; .
\label{zinf}
\end{eqnarray}
For $q\neq 0$ $f(z)$ is dominated by $a_D$, whose  
phase at infinity varies over the range $[0,\pi]$.
Results of a quantitative analysis using Mathematica
are shown in figure~6.
\begin{figure}[htb]
\centerline{\epsfxsize=2.4in\epsffile{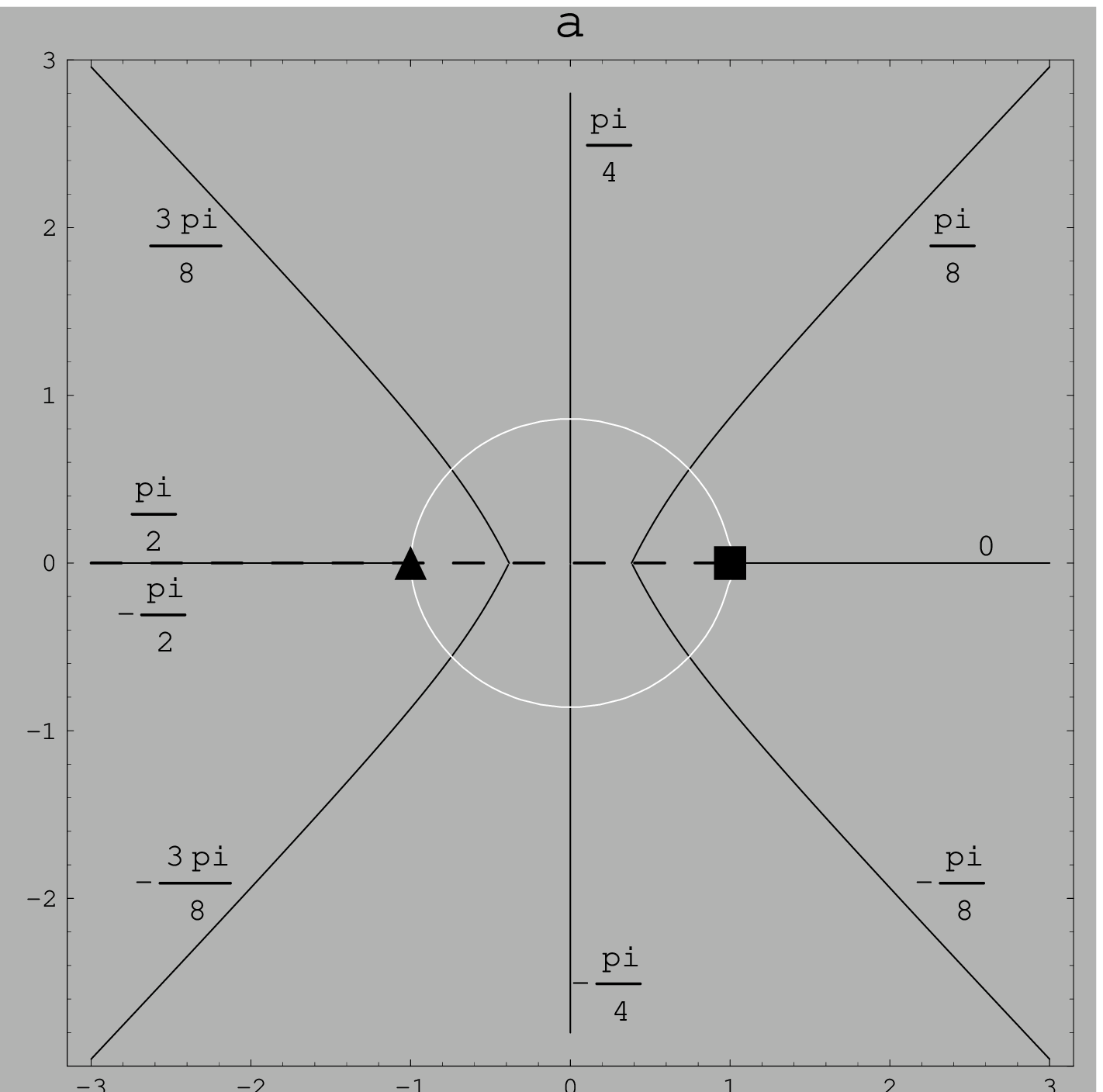}
\hspace{10 pt}
\epsfxsize=2.4in\epsffile{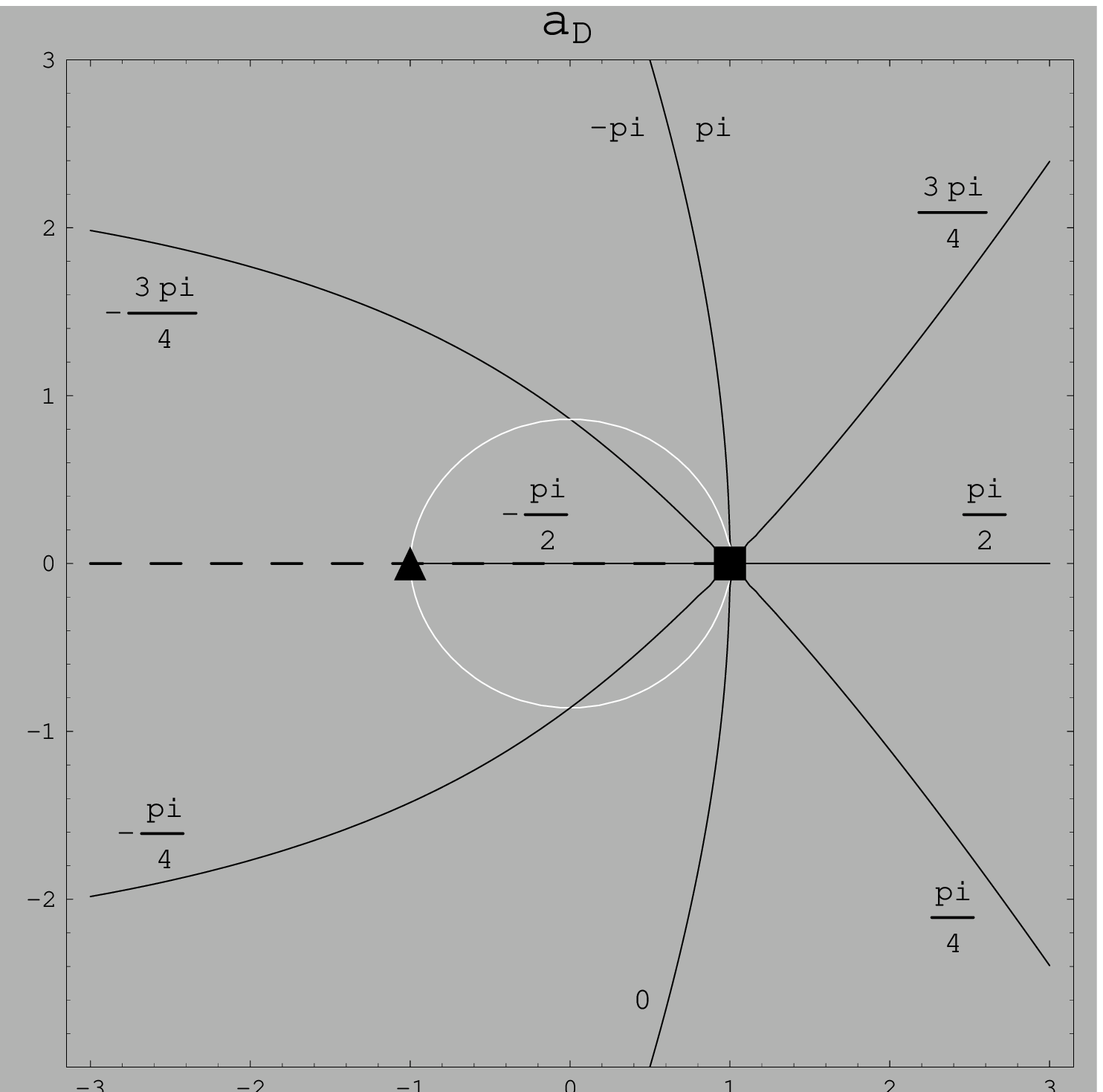}}
\vspace{10 pt}
\centerline{\epsfxsize=2.4in\epsffile{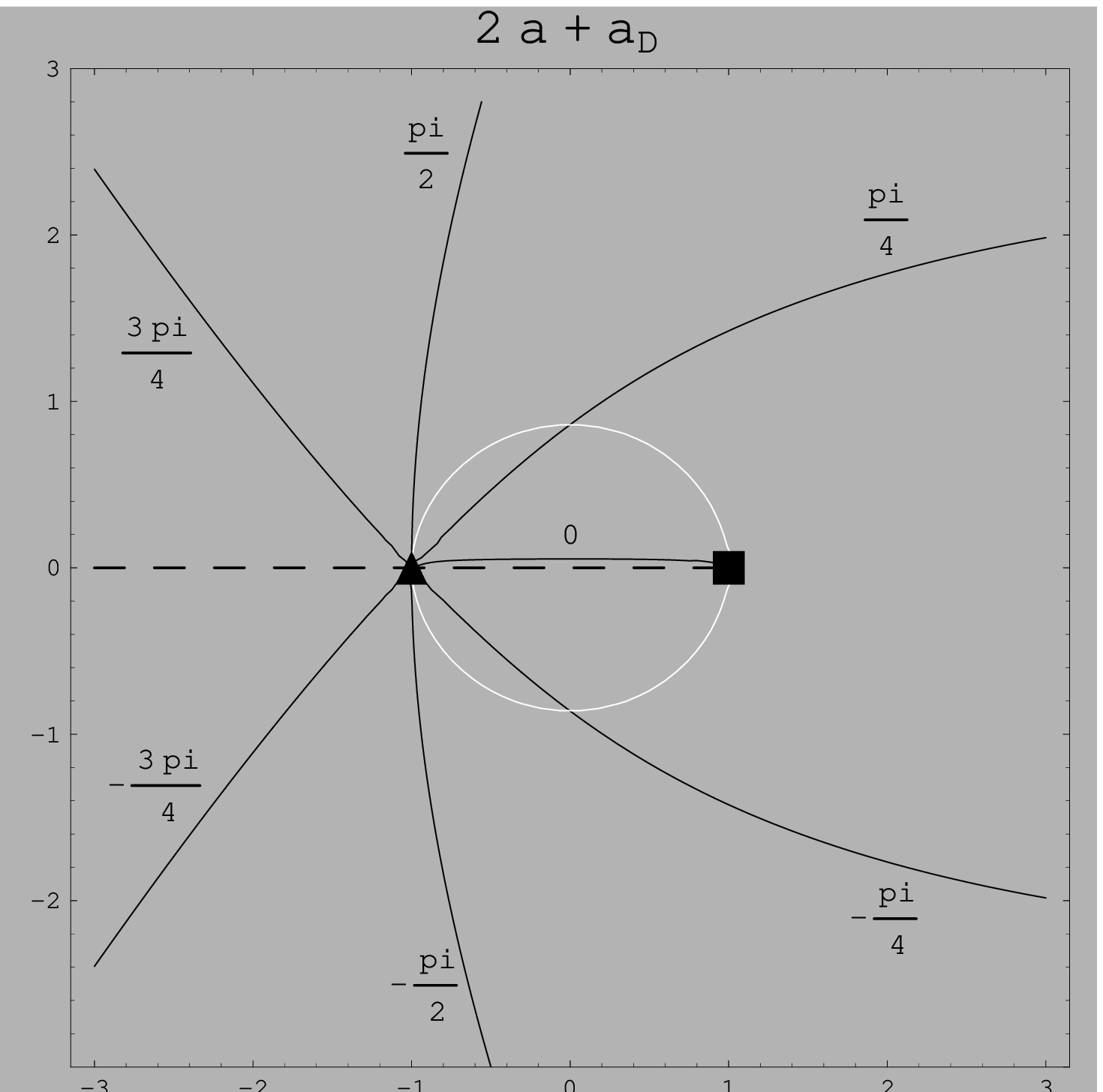}
\hspace{10 pt}
\epsfxsize=2.4in\epsffile{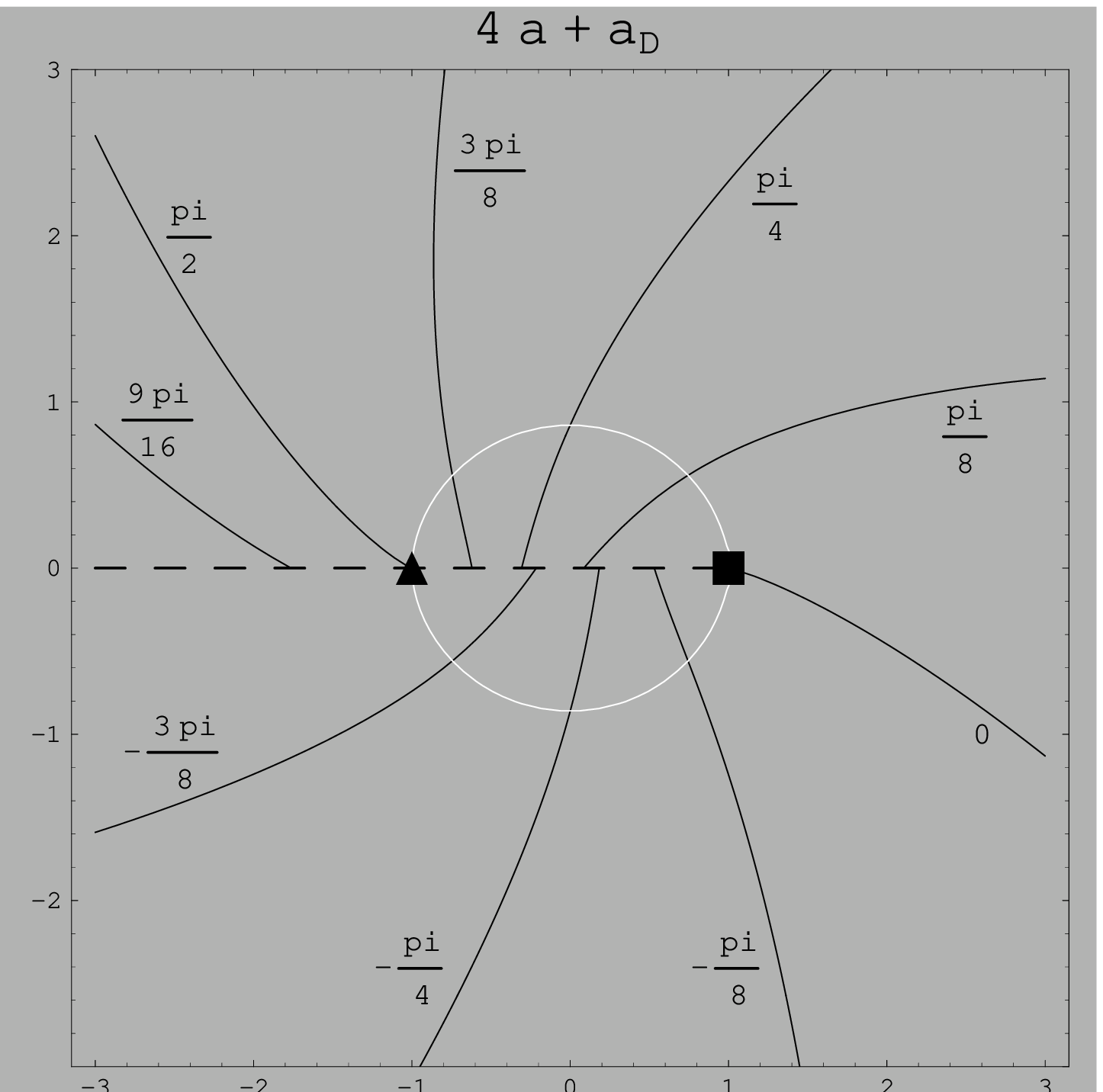}}
\caption{CPC's for $(1,0), (0,1), (2,\pm1)$, and $(4,1)$.
The sign in the third case is $(+)$ in the upper half
plane, and $(-)$ in the lower half plane.}
\end{figure}

\end{document}